\documentclass[conference]{IEEEtran}

\usepackage{graphicx}
\usepackage{url}
\usepackage{xspace}

\newcommand{\entk}{EnTK\xspace}
\newcommand{\rct}{RADICAL-Cybertools\xspace}
\newcommand{\rp}{RADICAL-Pilot\xspace}

\begin{document}

\title{Middleware Building Blocks for Workflow Systems}

\author{Matteo Turilli$^{1}$, 
	    Vivek Balasubramanian$^{1}$, 
	    Andre Merzky$^{1}$, 
	    Ioannis Paraskevakos$^{1}$,
	    Shantenu Jha$^{1}$$^{,2}$\\
	\small{\emph{$^{1}$ Rutgers, the State University of New Jersey, Piscataway, NJ 08854, USA}}\\
	\small{\emph{$^{2}$ Brookhaven National Laboratory, Upton, New York, 11973}}\\
}

\date{}
\maketitle

\begin{abstract} 
This paper describes a building blocks approach to the design of scientific workflow systems. We discuss RADICAL-Cybertools as one implementation of the building blocks concept, showing how they are designed and developed in accordance with this approach. This paper offers three main contributions: (i) showing the relevance of the design principles underlying the building blocks approach to support scientific workflows on high performance computing platforms; (ii) illustrating a set of building blocks that enable multiple points of integration, ``unifying'' conceptual reasoning across otherwise very different tools and systems; and (iii) case studies discussing how RADICAL-Cybertools are integrated with existing workflow, workload, and general purpose computing systems and used to develop domain-specific workflow systems.
\end{abstract}


\section{Introduction}\label{sec:intro}

Sophisticated and scalable workflows have come to epitomize advances in
computational science. To the credit of workflow systems initially developed
for ``big science'' projects, such as those in high-energy physics, many
advances were made when the scientific distributed computing infrastructure
and software ecosystem was missing important features and relatively fragile
when compared to today. Many successful workflow systems evolved to support
the end-to-end execution of workflows.

The landscape of scientific application requirements and software
infrastructure has changed. Although high-throughput execution of tasks---the
original driver of ``big science'' workflows---is still important, it is
joined by other functional and automation requirements. New application
scenarios involve the time-sensitive integration of experimental data from
large-scale instruments and observation systems with high-performance
computing. Workflows are also becoming more pervasive across application
types, scales and communities. Scientific insight typically requires
computational campaigns with multiple distinct workflows, heterogeneous tasks
and distinct runs. For example, an application may involve distinct phases of
parameter exploration and optimization, sensitivity analysis and uncertainty
quantification.

Previously missing software infrastructural capabilities that necessitated
the development of end-to-end workflow systems are now relatively more
reliable, better supported and more consistently available. The emergence of
diverse Python-based task distribution and coordination systems, Apache data
analysis tools, and container technologies provide useful examples.

An important and increasingly prevalent trend is that application developers
tend to develop their own workflow solutions, tailored to the requirements of
their applications. Ref.~\cite{workflow-systems-url} enumerates in excess of
230 purported workflow systems: some partial, others closer to being
end-to-end; some specific to a workload or functionality, others
general-purpose; some stand-alone, others designed to be integrated with
other systems.

The proliferation of workflow systems raises many questions for users and
developers. How to support the agile development and composition of workflow
systems that share capabilities, while not constraining functionality,
performance, or sustainability? How to lower the barrier for leveraging
existing software infrastructure? How to provide a sustainable ecosystem of
both existing and new software components from which tailored workflow
systems can be composed? These questions  are set against trends of
increasing functional requirements and sophistication of workflows.

This paper advocates a building blocks approach to the design and development
of scientific workflow systems. We postulate building blocks leverage
emerging trends in software and distributed computing infrastructure, and the
approach supports a sustainable ecosystem of both existing and new software
components from which tailored workflow systems can be composed. Building
blocks enable expert contributions while lowering the breadth of expertise
required of workflow system developers. They render obsolete a focus on
developing a workflow system that purports to be ``better'' than other
workflow systems, and emphasizes if not incentivizes the community towards
development of collective capabilities.

After a brief description of the building blocks approach and its four design
principles of self-sufficiency, interoperability, composability, and
extensibility, Sec.~\ref{sec:rct} discusses how we used the building blocks
approach to develop \rct{} to enable the execution of workflows from diverse
scientific domains on High Performance Computing (HPC) platforms. These are a
set of software systems that can be used independently and integrated into
middleware, among themselves and with third-party systems. We introduce a
four-layered view of high-performance and distributed systems and we describe
how each system implements distinctive functionalities for each layer.

Sec.~\ref{sec:casestudies} discusses how \rct{} complement and contribute to
existing workflow systems and middleware. We present three case studies
integrating \rct{} with end-to-end workflow systems (Swift), workload
management systems (PanDA) and general purpose computing frameworks (Spark
and Hadoop), and a case study discussing the development of domain-specific
workflow systems (ExTASY, RepEx, HTBAC and ICEBERG). These case studies have
enabled diverse scientific applications, involving high-throughput computing,
multi-protocol simulations, adaptive execution, data-intensive simulations,
and image processing.

We conclude with a discussion of the practical impact of the case studies as
well as the lessons learned by testing the validity and feasibility of the
building blocks approach. We highlight the benefits of implementing new
capabilities into existing workflow systems by integrating the \rct{}. We
also outline the limitations of our contributions as well as some open
questions.

\section{Related Work}\label{sec:related}

We classify existing workflow systems into three categories, focusing only on
those with the highest adoption and ongoing development. All-inclusive
workflow systems such as Kepler, Swift, Fireworks, and Pegasus that provide
full-featured, end-to-end capabilities that include application creation,
execution, monitoring and provenance. General-purpose workflow systems such
as Ruffes, COSMOS, and GXPMake that enable end-to-end execution but
prioritize the simplicity of their interfaces, limiting the range of
capabilities. Finally, domain-specific workflow systems such as Galaxy,
Taverna, BioPipe, and Copernicus that provide interfaces tailored to the
requirements of specific domain scientists.

The decomposition of workflow systems into systems with high cohesion and low
dependency supports decoupling of independent software development efforts
and promotes the use of standardized interfaces. These systems are
implemented in monolithic or modular fashion to support specific
capabilities, and have been used to develop multiple workflow systems by
integration. For example, Spark, Hadoop, and MapReduce can be
integrated---with or without pipelining tools like Luigi, Toil, Airflow,
Azkaban or Oozie---to create special-purpose workflows
systems~\cite{vivian2017toil,zhang2015scientific}. Nonetheless, these tools
are specifically tailored to data-oriented workflows, face several
performance bottlenecks when ported to high-performance computing (HPC)
machines, and require dedicated deployment~\cite{chaimov2016scaling}.
Research in interoperability of HPC systems with data-parallel frameworks is
ongoing and provides and extends middleware to efficiently support
data-oriented workflows on HPC\@. A few examples are Pilot-Hadoop and
Pilot-Spark, Twister or Pilot-Streaming.

Modularity, in software deployment, has evolved from \texttt{chroot},
\texttt{jails} and Solaris \texttt{zones} and, more in general, to what is
called the ``UNIX Philosophy'' into modern day service oriented architecture
(SOA) and its Microservice variants~\cite{dragoni2017microservices}. These
approaches evolve from the concepts of Component Based Software
Engineering~\cite{heineman2001component} (CBSE) where computational and
compositional elements are explicitly
separated~\cite{batory1992design,garlan1995architectural,schneider2000components}.

We build upon CBSE and SOA concepts, investigating modularity at the level of
stand-alone software systems and not at the level of modules or routines of a
single system. In this context, we underline the benefits of CBSE-like
concepts when applied to workflow systems for scientific computing executed
on HPC resources. AirFlow, Oozie, Azkaban, Spark Streaming, Storm, or Kafka
are examples of tools that have a design consistent with the proposed
approach. Different from the CBSE and SOA approaches, we make the internal
states and events of each module accessible and we employ connectors and
translation layers between interfaces.

\section{building blocks Approach}\label{sec:bba}

Each building block has a set of entities, a set of functionalities that
operate on these entities, and a set of states, events and errors for each
entity. Architecturally, the building blocks design requires: (i) a
well-defined and stable interface for input and output that enables clean
separation between computational and compositional features; (ii) one or more
conversion layers capable of translating across diverse representations of
the same type of entity; (iii) one or more modules that implement the
functionalities to operate on these entities and expose higher-level
abstractions for their composition.

In our adaptation, the building blocks approach is based on four design
principles: self-sufficiency, interoperability, composability, and
extensibility. Self-sufficiency and interoperability depend upon the choice
of both entities and functionalities. Entities have to be general enough so
that specific instances of that type of entity can be reduced to a unique
abstract representation. Accordingly, the scope of the functionalities of
each building block has to be limited exclusively to its entities. In this
way, interfaces can be designed to receive and send diverse codifications of
the same type of entity, while functionalities can be codified to
consistently translate those representations and operate on them.

Composability depends on whether the interfaces of each building block
enables communication and coordination. Blocks communicate information about
the states, events and errors of their entities, enabling the coordination of
their functionalities. Due to the requirement of self-sufficiency, the
coordination among blocks cannot be assumed to happen implicitly but has to
be codified on the base of an explicit model of the entities' states. The
sets of functionalities of a block need to be extensible to enable the
coordination among states of multiple and diverse blocks. Note that
extensibility remains bound by both interoperability and self-sufficiency.

Each design principle of the building blocks approach poses unique challenges
when applied to software systems that can be used both standalone and
integrated with third-party systems. Choosing entities and scoping
functionalities to enable self-sufficiency requires expanding the design
phase and therefore longer development iterations. Further, interoperability
requires system-level interfaces to become a first order concern and to be
based on well-defined, general purpose abstractions. The coordination
protocols that enable composability require generalization of variable
access, dataflow and procedure calls. Extensibility also requires shared
coding convention and documentation.

The building blocks approach does not reinvent modularity, it applies it at
system level to enable composability among independent software systems. As
an abstraction, modularity enables separation of concerns by encapsulating
discrete functions into semantic units exposed via a dedicated interface. As
such, modularity can be used both at function and system level. Modularity at
function level depends on the programming paradigm and the facilities offered
by programming languages. Modularity at system level depends on the interface
exposed by each system.

Traditionally, components of software systems independently designed by third
party organizations have been difficult to integrate outside the well-defined
scope of an operating system like, for example, Unix. While interfaces can
hide implementation details, working as implementation-independent
specifications of capabilities, integration still requires semantic
uniformity across interfaces. Obtaining such uniformity is challenging and
largely unsupported by specific constructs both at specification and
programming level. Further, integrating independent systems poses challenges
in language heterogeneity, error handling, input/output validation, effective
documentation and comprehensive testing.

Building blocks approach contributes to address integration challenges across
independent systems by specifying state, event and error models for each
block. Following best practices in application program interface design,
entities are explicitly specified and implemented in the block's interface
and used as input for each exposed functionality. Each entity has a set of
associated states, events and errors. The order of the state is guaranteed by
the implementation (e.g., a task cannot be executed before being scheduled
and scheduled before being bound to a resource) while events are unordered
but always contained within two defined states. Errors are always associated
to an entity, state and event. Communication is decoupled from coordination
and independent from the implementation of communication channels.

Even when applied at system level, modularity, and therefore the building
blocks approach, presents at least two major trade offs. Building systems as
blocks increases design and implementation effort, making unfeasible an
unstructured but rapid development approach. While unstructured approaches
are counter productive for long-term maintenance, short-term solutions would
pay an unpractical overhead to the building blocks approach. Further,
integration of systems that are independently developed imposes sharing
responsibility of software reliability across multiple stakeholders. Often,
this can be undesirable as user attributes all the responsibility to the
stakeholder of their immediate interface. This problem can be mitigated by
system-level fault tolerance but it remains an element to carefully evaluate
when considering the building blocks approach.

\section{RADICAL-Cybertools}\label{sec:rct}

\begin{figure}[t]
  \includegraphics[width=\columnwidth]{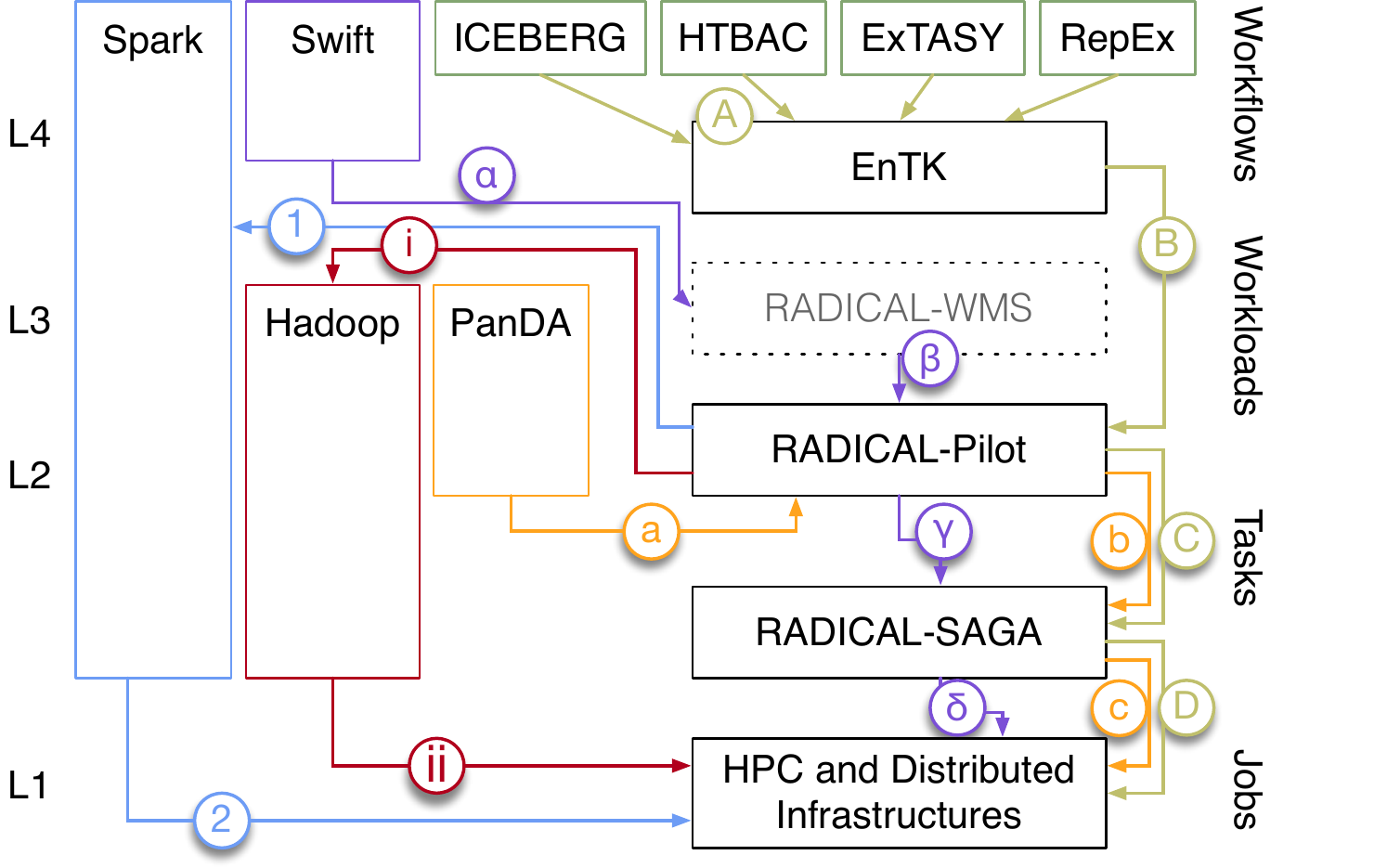}
  \caption{Composition of RADICAL-Cybertools (black) with domain specific
           workflow systems (green, A--D), workflow system (purple,
           $\alpha$--$\delta$), workload management system (orange, a-c),
           framework for distributed data processing (red, i-ii), and a
           unified analytics engine (blue, 1--2). Numbered layers on the
           left; names of entities on the right. Solid colored lines indicate
           various integrations points with \rct{}; Dashed boxes indicate
           tools still under development.}\label{fig:radical-bb}
\end{figure}

\rct{} are software systems designed and implemented in accordance with the
building blocks approach. Each system is independently designed with
well-defined entities, functionalities, states, events and errors.
Fig.~\ref{fig:radical-bb} shows three existing \rct{} systems: RADICAL
Ensemble Toolkit (hereafter simply referred to as \entk{}), RADICAL-Pilot and
RADICAL-SAGA\@. RADICAL-WMS is a workload management system (WMS) still under
development.

\begin{figure}[!ht]
  \includegraphics[width=\columnwidth]{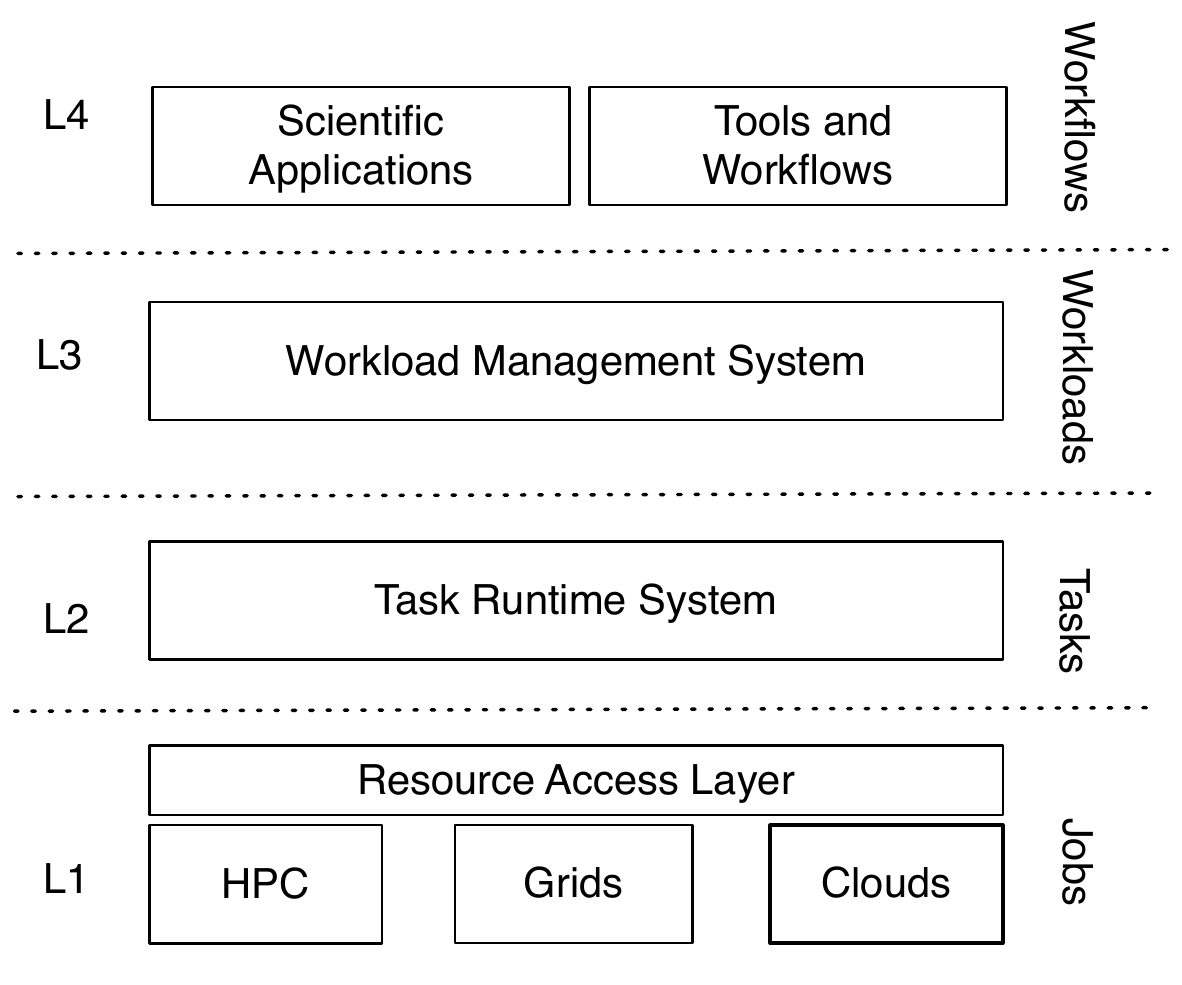}
  \caption{Primary functional levels. The diagram supports an analysis of the
           functional requirements for workflow systems, and the
           primary entities at each level, agnostic of the applications and
           resources.}\label{fig:layers}
\end{figure}

Individual \rct{} are designed to be consistent with a four-layered view of
distributed systems for the execution of scientific workloads and workflows
on HPC resources. Each layer has a well-defined functionality and an
associated ``entity''. The entities are \textbf{workflows} (or applications)
at the top layer and resource specific \textbf{jobs} at the bottom layer,
with \textbf{workloads} and \textbf{tasks} as intervening transitional
entities in the middle layers. The diagram of Fig.~\ref{fig:layers} provides
a reference example for the integration among entities across layers that is
independent of the specifics of applications,
\rct{} and resources.

\textit{Workflow and Application Description Level (L4)}: Requirements and
semantics of an application described in terms of a workflow.

\textit{Workload Management Level (L3):} Applications devoid of semantic
context are expressed as workloads which are a set of tasks that can be
executed concurrently. The Workload Management layer is responsible for: (i)
the selection and configuration of available resources for the given
workload; (ii) partitioning the workload over the selection of suitable
resources; (iii) binding of tasks to resources.

\textit{Task Execution Runtime Level (L2):} L3 delivers tasks to L2 which is
responsible for their execution on the selected resources. L2 is a passive
recipient of tasks from L3 but includes functionalities to acquire the
indicated HPC resources, schedule the given tasks over available resources,
and execute these tasks with the indicated data and number of cores.

\textit{Resource Layer (L1):} The resources used to execute tasks are
characterized by their capabilities, availability and interfaces. Different
resources present inconsistency in the way capabilities are provisioned but
advances in syntactically uniform resource access layers enable task
execution across resources.

Currently, in \rct{} each task defines an executable, e.g., \texttt{python},
\texttt{GROMACS}, \texttt{AMBER}, \texttt{SPECFEM} or any other executable
programs. Each task description contains the arguments to pass to the
executable, the type of parallelism required (e.g., message passing interface
(MPI), open multi-processing (OpenMP)), the type and number of processing
unit (CPU or GPU), the amount of memory, and the data staging requirements.
\rct{} implement full task isolation, enabling the concurrent or sequential
execution of heterogeneous, dependent or independent executables on a given
set of acquired resources. \rct{} are agnostic of the operations performed by
each executable: for each task, \rct{} satisfy its dependences, set up its
environment and spawn its execution waiting for the executable to return in a
final state. Therefore, \rct{} do not have access to the operations performed
by each executable.

\rct{} conform to the principles of self-sufficiency, interoperability,
composability and extensibility. \entk{} exposes an application programming
interface (API) for the description of scientific applications as static or
dynamic sets or sequences of pipelines. Pipelines are sequences of stages
that, in turn, are sets of tasks. Sequences and sets formally define the
relationship of priority among task executions: the tasks of a stage execute
concurrently, tasks of different stages of the same pipeline execute
sequentially, and pipelines execute concurrently. Resources are acquired and
managed via a third-party runtime system that executes tasks on the acquired
resources.

\entk{} is self-sufficient as it enables necessary and sufficient
functionalities for its set of entities, independently from third-party
software systems. \entk{} is interoperable because different representations
of a workflow (e.g., directed acyclic graph (DAG)) can be converted to
pipelines of stages of tasks, and because it is agnostic towards runtime
systems and the type of resources on which they execute tasks. \entk{} is
also composable because it enables arbitrary coordination protocols (e.g.,
push/pull or master/worker) by explicitly defining the state model of its
entities. Finally, \entk{} is also extensible as new capabilities can be
implemented for its entities, e.g., adaptivity of both workflows structure
and task specifications at runtime.

RADICAL-Pilot is a pilot system that exposes an API to enable the acquisition
of resources on which to schedule tasks for execution. The design of
RADICAL-Pilot includes pilot and compute unit as entities. Capabilities are
made available to describe, schedule, manage and execute entities. Pilots,
units and their functionalities abstract the specificities of diverse types
of resource, enabling the use of pilots mainly on single and multiple HPC
machines, but also on high-throughput computing (HTC) and cloud
infrastructures. A pilot can span single or multiple compute nodes, resource
pools, or virtual machines. Units of various size and duration can be
executed, supporting MPI and non-MPI executables, with a wide range of
execution environment requirements.

The design of RADICAL-Pilot is: self-sufficient because, as with \entk{}, it
independently implements the necessary and sufficient set of functionalities
for its entities; interoperable in terms of type of task, resource, and
execution paradigm; and extensible as new properties can be added to the
pilot, unit and resource descriptions, and more functionalities can be
implemented for these entities. Currently, composability is partially
designed and implemented: while the API can be used by both users and other
systems to describe generic tasks for execution, RADICAL-Pilot requires
RADICAL-SAGA to interface to HPC resources. A prototype interface to cloud
resources based on LibCloud is available and a general-purpose resource
connector component is under development.

RADICAL-SAGA exposes a homogeneous programming interface to the queuing
systems of HPC resources. SAGA---an Open Grid Forum (OGF)
standard---abstracts away the specificity of each queue system, offering a
consistent representation of jobs and of the capabilities required to submit
them to the resources. The design of RADICAL-SAGA is based on the job entity
and its functionalities enable job submission and jobs' requirements handling
(self-sufficiency). Both entities and functionalities can be extended to
support, for example, new queue systems or new type of jobs (extensibility).
The SAGA API resolves the differences of each queue system into a general and
sufficient representation (interoperability), exposing a stable set of
capabilities to both users and/or other software elements (composability).

Currently, data staging capabilities are implemented in each \rct{} via
third-party tools like SFTP, SCP, and Globus Online. Nonetheless, we do not
have a dedicated cybertool for managing data storage, provenance and
archiving. Users can enable these capabilities by integrating third-party
systems into \entk{} workflows and RADICAL-Pilot workloads, creating their
own data management steps. Integration of third party systems is facilitated
by implementing task and compute unit in \rct{} as wrappers for
self-contained programs. Data management tools can be executed or accessed
independent from the coordination, communication and runtime environment
requirements of \rct{}.

\section{Building Blocks, \rct{} and Workflows
Systems}\label{sec:casestudies}

\rct{} as a whole are not an end-to-end workflow system. Each cybertool is an
independent system that can also be integrated with other systems (\rct{} or
otherwise) to form tailored middleware solutions. For example, several
independent communities directly utilize RADICAL-SAGA alone, with
RADICAL-Pilot or other pilot systems like, for example, PanDA Pilot. Other
communities integrate all \rct{} with or without third-party systems to
support the execution of diverse types of scientific workflows. Thus, \rct{}
are not posed to replace existing workflow systems: \rct{}' novelty is to
enable the integration across systems independently developed and not
necessarily designed to integrate. Crucially, this includes existing
workflow, workload and computing frameworks, alongside their components.

We believe an ecosystem in which end-to-end workflow systems and building
blocks coexist and, when useful, are integrated helps to avoid both lock-in
and fragmentation. Such an ecosystem would allow scientists with specific and
stable requirements to use an end-to-end system while others to aggregate
existing capabilities into tailored solutions. Inversely, non-integrable
systems built with slightly different capabilities fragments the user
experience, forcing scientists to learn to use multiple systems, depending on
the context in which they have to operate.

As building blocks, \rct{} offer several benefits when used to describe and
execute scientific workflows. Among these benefits, the most relevant is
isolating scientists from job management (L1), task management (L2), and
workload management (L3). These capabilities are further abstracted away in
L4, letting scientists to exclusively focus on workflow description and
application logic. Note that while this isolation is offered by other
systems, \rct{} is agnostic towards which software tools and systems are
integrated at each layer L1--4.

When integrated, \rct{} simplify the codification of workflows, lowering the
barrier to adoption, maintenance and reuse. When using \entk{}, workflows are
codified as pipelines in a general purpose language (Python) and
application-specific constructs (Task, Stage, Pipeline and AppManager). As
programs, workflows can be maintained following diverse approaches: from
keeping a simple script on a scientist's workstation to sharing a more
complex application among multiple scientists via a collaborative version
control system. Codifying workflows as code but without a dedicated domain
language, offers the opportunity of reusing a portion of code in the form of
methods, classes and modules. Further, scientists have the option to grow the
code as needed, typically starting from a small script and growing it into an
application as the research advances, alone or with the help of other
scientists and software engineers.

Interoperable, extensible and logically self-contained software blocks,
alongside lower technical barriers to their composability allow designing
workflows as domain-specific applications. These type of applications solves
classes of scientific problems, not issues of resource and execution
management. Domain-specific applications, alongside the blocks they use,
become sustainable because they can be understood and maintained by diverse,
invested communities. This is the sustainability model of successful open
source software, including some of the existing solutions for certain types
of workflows and resources, e.g., the Apache Hadoop ecosystem.

Supporting the development and maintenance of domain-specific applications is
becoming increasingly important to enable scientific workflows. Alongside
large communities in which the same workflow is used for many years (e.g.,
the LHC community), many research fields increasingly require running
rapidly-evolving workflows with relatively short computation campaigns. These
workflows depend on simulations and analysis procedures that evolve during
the campaign, integrating new models and methodologies. As such they require
a software ecosystem with independent systems that can be easily integrated
and extended, depending on evolving scientific requirements.

\subsection{Integrating End-to-end Workflow Systems}\label{ssec:end-to-end}

End-to-end workflow systems enable a wide range of capabilities on several
types of computing resources. Often, their adoption and deployment require
investing sizable amount of resources, developing system-specific code and
establishing dedicated processes. Extending this type of workflow systems
requires advanced development knowledge both at system and resource level,
and taking into account the requirements of a widely used and
production-grade code base. Integrating building blocks with these end-to-end
system may lower the amount of resources needed to implement new
functionalities, while requiring moderate development skills.

As an example of how the building blocks approach can be utilized in other
systems, we map the primary functional levels described in
Fig.~\ref{fig:layers} to Pegasus~\cite{Deelman2005}, one of the most adopted
end-to-end workflow system. Scientific applications are described as abstract
workflows using the HubZero API, or workflow composition tools such as Wings
and Airvata. These interfaces correspond to the application layer (L4) of
Fig.~\ref{fig:layers}.

The abstract workflow is transformed to a concrete workflow by the Mapper
component. The transformation takes into account the availability of
software, data, and computational resources required for execution, and can
restructure the workflow to optimize performance. A concrete workflow with
several interdependent jobs, each consisting of several interdependent tasks,
is passed to a workflow engine. Pegasus utilizes different engines, depending
on the target resource: (1) lightweight execution engine for local resources;
(2) HTCondor DAGMan and HTCondor Schedd for clusters and HPC platforms; and
(3) HTCondor with Glide-in WMS for grids. Functionally, Mapper, workflow
engine, and local scheduler correspond together to the workload management
layer (L3) of Fig.~\ref{fig:layers}.

Pegasus supports three  modes of job execution, depending on the execution
environment and architecture of the remote machine: (1) PegasusCluster, a
single-threaded engine that submits one task at a time; (2) PegasusLite, for
handling tasks input and output data on resources with no shared filesystem;
and (3) Pegasus MPICluster, for systems with a shared filesystem where MPI is
used to implement a master-slave layout for task binding and execution.
Collectively, these three remote execution engines correspond to the task
runtime layer (L2) of Fig.~\ref{fig:layers}.

Pegasus uses GlobusGRAM and CREAM-CE to submit jobs directly to remote
batch-queuing systems and resource managers such as Simple Linux Utility for
Resource Management (SLURM), Portable Batch System (PBS), Platform Load
Sharing Facility (LSF), and Sun Grid Engine (SGE). These tools correspond to
the resource access layer (L1) of Fig.~\ref{fig:layers}.

Following this mapping, end-to-end workflows systems or some of their
component can be integrated with \rct{} as building blocks to enable new
capabilities. Together with the Swift development team, we used this approach
to integrate Swift~\cite{wilde2011swift} with RADICAL-Pilot and RADICAL-SAGA.
Swift has a long development history, with several versions that supported
diverse case studies. Swift also integrated pilot systems of which Coasters
is actively supported. The design of Swift is modular and it relies on
connectors to interface with third-party systems.

In Swift, the language interpreter and the workflow engine are tightly
coupled but connectors can be developed to stream the tasks of workflows to
other systems for their execution. As seen in Sec.~\ref{sec:rct},
RADICAL-Pilot can get streams of tasks as an input and submit these tasks to
pilots for execution.

We integrated Swift with RADICAL-Pilot to enable the distributed and
concurrent execution of Swift workflows on multiple HPC platforms and HTC
infrastructures (Fig.~\ref{fig:radical-bb}, purple $\alpha$--$\delta$). The
distributed scheduling capabilities of RADICAL-Pilot offered the possibility
to minimize the time to completion of tasks execution, obtaining both
qualitative and quantitative improvements~\cite{turilli2017evaluating}.
Qualitatively, \rct{} enabled Swift to execute workflows concurrently on both
HPC and HTC resources via late binding of both tasks to pilots and pilots to
resources. Quantitatively, the time to completion of workflows was improved
by leveraging the shortest queue time among all the target resources.

\begin{figure}
    \centering
    \includegraphics[width=0.49\textwidth]{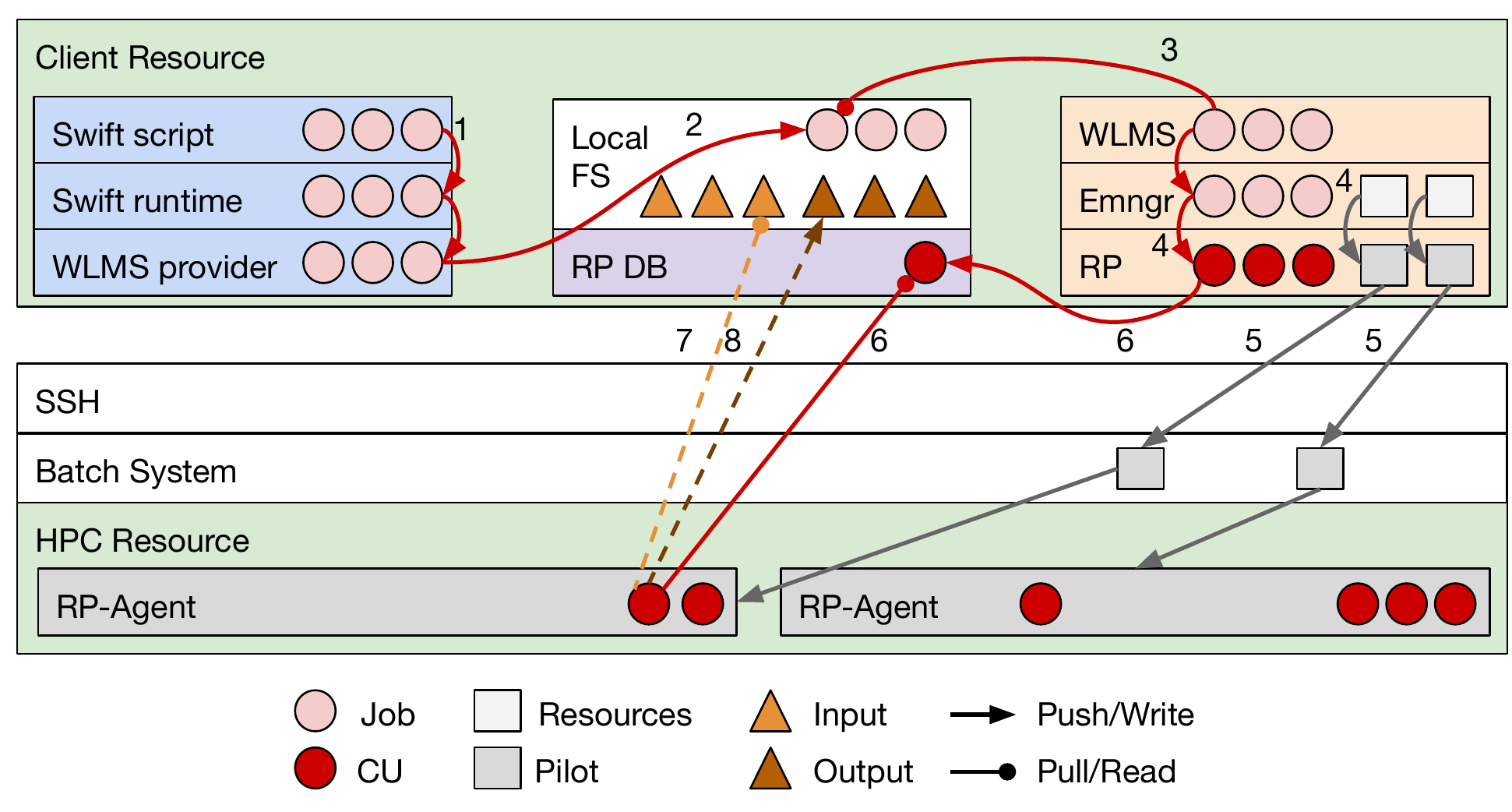}
    \caption{Integration between Swift and RADICAL-Pilot. The two systems
             exchange task descriptions via a local filesystem. RADICAL-WMS
             derives the size and duration of the pilots from the task
             requirements, independently from
             Swift.\label{fig:swift_integration}}
\end{figure}

The integration with RADICAL-Pilot required the Swift team to develop a
dedicated connector by iterating on the already available shell connector
(Fig.~\ref{fig:swift_integration}). We used the opportunity to prototype a
distributed workload management (RADICAL-WMS) as a means of research. The
connector enabled saving task descriptions on the local filesystem from where
RADICAL-WMS was able to load and parse these descriptions without needing any
added functionality. This type of integration was made possible by sharing
the task entity semantics between the two systems and by isolating distinct
functionalities operating on that entity in two distinct software systems.
Note how these two systems were not designed to be integrated and were
developed by independent teams.

\subsection{Integrating a Workload Management System}\label{ssec:panda}

Often, workflow managers are developed to support specific resources,
workloads, projects and communities. Extending their capabilities can be
difficult, especially when the new capabilities do not serve the intended
core use cases. Instead of developing yet another domain-specific workload
manager, integration with existing building blocks can represent an economic
and viable solution. This was true for PanDA, a workload management system
designed to support execution of independent tasks on Grid computing
infrastructures like WLCG~\cite{maeno2008panda} and, to a lesser extend,
leadership-class HPC platforms.

Executing large number of small jobs on leadership HPC platforms presents two
main challenges: using a queue system that privileges large MPI jobs; and
accessing untapped resources without disrupting the overall utilization of
the machine. Pilots can address both challenges but pilot systems are
difficult to deploy on HPCs. The main problem is efficiently managing the
concurrent and sequential execution of small heterogeneous jobs at scale.

We developed an interface to RADICAL-Pilot called Next Generation Executer
(NGE). NGE enables workload management systems designed for HTC to execute
workloads on pilots instantiated on leadership-class HPC platforms
(Fig.~\ref{fig:radical-bb}, orange a--c). As part of their workload
management system, the PanDA team developed Harvester, a job broker to
support the execution of part of the ATLAS Monte Carlo workflow on Titan.
Harvester misses pilot capabilities and the PanDA team developed an Harvester
connector to NGE instead of implementing a new pilot system. In this way,
event simulations of the ATLAS project can be executed both concurrently and
sequentially on the resources acquired by submitting a single job to Titan's
queue.

\begin{figure}
    \centering
    \includegraphics[width=0.49\textwidth]{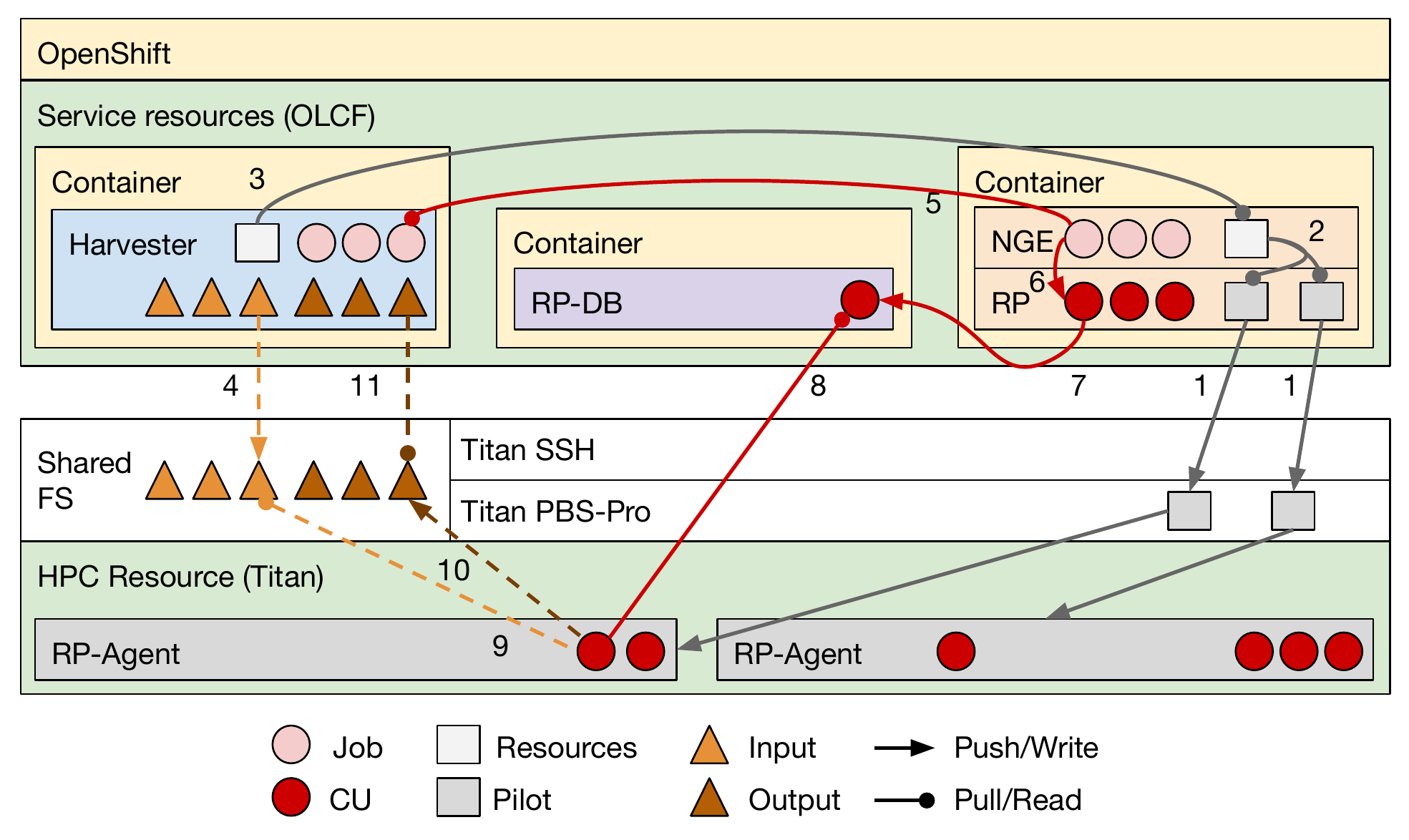}
    \caption{Integration between PanDA and RADICAL-Pilot via the Next
             Generation Executer (NGE) REST interface. All systems execute on
             OLCF service resources within containers. Pilots are exposed to
             PanDA as an aggregation of available resources (steps 2 and
             3).\label{fig:panda_integration}}
\end{figure}

Harvester uses NGE to exchange information about tasks descriptions and
resources requirements, while RADICAL-Pilot behaves like a resource queue for
Harvester (Fig.~\ref{fig:panda_integration}). Both systems require no
modifications to be integrated but the development of a coordination protocol
to pull/push information about entities and their states. As with Swift,
PanDA Broker and RADICAL-Pilot are independently developed and their
integration was performed when the two stacks were already in production.

\subsection{Integrating General Purpose Computing
Frameworks}\label{ssec:apache}

Computing frameworks like Hadoop and Spark offer specific capabilities,
programming models and a large ecosystem of related software modules. As with
end-to-end workflow systems, users tend to produce code and processes that
depend on these frameworks that need to be maintained while scaling and
runtime requirements evolve. Deploying and using these frameworks on HPC
resources is challenging, especially when considering machines that are not
built and configured as Hadoop or Spark clusters. Integration with the \rct{}
building blocks addresses these challenges and offer the same multi-task
programming interface across both frameworks.

Hadoop exposes an API for users and other software systems (composability) to
describe distributed applications, mostly in terms of the MapReduce
programming model, supporting distributed filesystem capabilities.
Accordingly, Hadoop implements the necessary and sufficient functionalities
of a workload management system (self-sufficiency). Hadoop can aggregate and
manage diverse storage resources via a master/worker subsystem composed of
multiple Namenode and DataNode instances (interoperability), and supports
diverse runtime systems like Mesos, YARN, and others, to schedule and execute
tasks on computing resources (extensibility).

Spark can be considered as a self-sufficient implementation of a Workflow
system. It enables necessary and sufficient workflow functionalities, for
machine learning, iterative analytics and streaming, independent of the
underlying Task Runtime System. Spark enables interoperability by supporting
different execution engines, such as Hadoop, MPI and others. Spark exposes an
API that can be used to develop distributed applications (composability).
Spark can be extended to support different types of workflows.

We integrated Hadoop, Spark and \rp{} into a framework dedicated to HPC
machines~\cite{hadoop-on-hpc} (Fig.~\ref{fig:radical-bb} red i--ii and blue
1--2). The integration addresses one of the major problems of using Hadoop
and Spark on HPC resources, avoiding the need for dedicated deployment and
customizations while retaining the full functionalities of both systems.
\rp{} configures, starts and manages a Hadoop/Spark cluster, and then
executes a user's Hadoop/Spark application on that cluster. Once done, \rp{}
shuts down the cluster and cleans up the environment.

We used \rp{}'s integration with Spark to parallelize MDAnalysis and
characterize its performance. MDAnalysis is a Python library that provides a
comprehensive environment for filtering, transforming and analyzing molecular
dynamics simulation trajectories~\cite{michaud2011mdanalysis}. Currently, we
use \rp{}'s integration with Spark to support diverse imagery analysis
algorithms for geological and polar sciences.

\subsection{Domain Specific Workflow Systems}\label{ssec:dsw}

We call a workflow system that provides a specific higher-level functionality
a Domain Specific Workflow system (DSW). We aggregated \rct{}, developing
four DSW: ExTASY, RepEx, HTBAC and ICEBERG (Fig.~\ref{fig:radical-bb}, green
A--D). Driven by specific application needs, each DSW is characterized by a
unique execution and coordination pattern and can serve multiple
applications.

\begin{figure}
    \centering
    \includegraphics[width=0.49\textwidth]{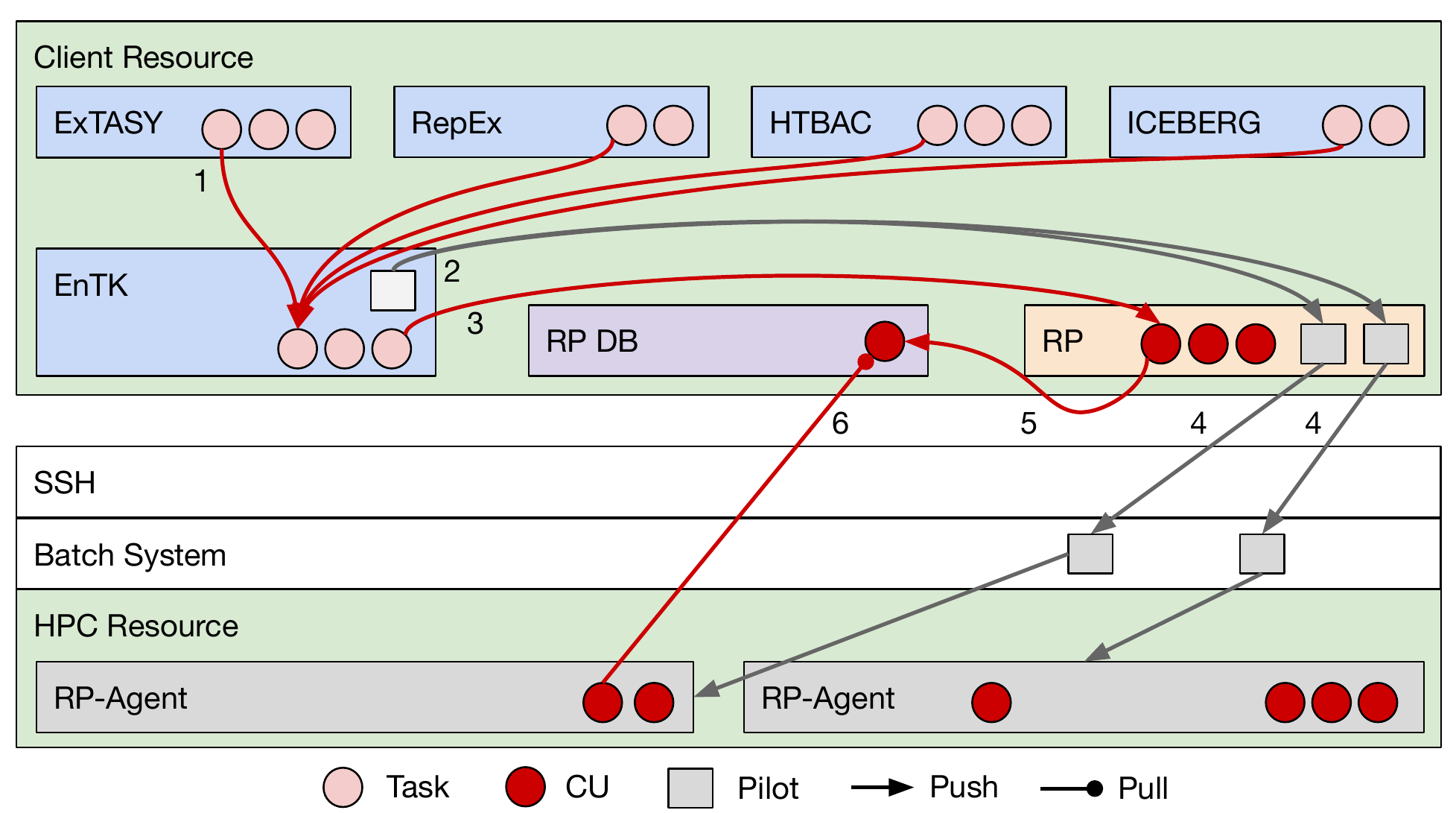}
    \caption{Integration between four domain-specific workflow (DSW)
            systems---ExTASY, RepEx, HTBAC, ICEBERG---and \entk{}. Numbers
            indicate the execution flow. RADICAL-Pilot (RP) database (DB) can
            be deployed on any host reachable from the
            resources.}\label{fig:entk_integration}
    \vspace*{-1em}
\end{figure}

Our DSW systems use \entk{} to support ensemble-based workflows. \entk{} is
agnostic to the details of the specific executables run by the ensemble
members, and the system used to manage their execution.
Fig.~\ref{fig:entk_integration} shows how \entk{} couples with RADICAL-Pilot
to execute the ensembles via pilots on HPC resources. Note that, in
principle, \entk{} could use a different runtime system and type of computing
infrastructure.

ExTASY provides the simulation-analysis execution pattern to support several
biomolecular sampling methods like DM-d-MD and CoCo~\cite{extasy}. RepEx
enables multiple replica-exchange methods which vary in the coordination
patterns across replicas, e.g., global synchronization barrier, or pair-wise
synchronization etc. RepEx supports synchronous and asynchronous
multi-dimensional exchange schemes~\cite{ct500776j}, separating performance
and functional layers while providing simple methods to extend interfaces.
HTBAC implements multiple pipelines of heterogeneous tasks for binding free
energy calculations, wherein both pipelines and tasks within a pipeline can
change at runtime. All three biomolecular DSW run on several HPC platforms,
including Oak Ridge National Laboratory (ORNL)'s and National Center for
Supercomputing Applications (NCSA)'s leadership-class machines. ICEBERG
supports scalable image analysis applications using multiple concurrent
pipelines.

ExTASY, RepEx, HTBAC and ICEBERG benefit from integrating \rct{} by not
having to reimplement workflow processing, task management and task execution
capabilities on distinct and heterogeneous platforms. This, in turn, enables
users to focus on customizing each DSW based on the requirements of specific
scientific domains. DSW and \rct{} free scientists from developing
capabilities outside their domain of expertise and from dealing with
resource- and middleware-specific deployment issues.

\section{Discussion and Conclusions}\label{sec:discussion}

Traditionally, assumptions about types of applications or resources have led
to software systems that, while modular, have not allowed reuse outside their
original requirements. We believe this is why functionalities pertaining to
entities like tasks or pilots are often reimplemented. Each system serves
well the single research group or the large scientific project but not each
other.

As argued in Sec.~\ref{sec:bba}, building blocks---i.e., self-sufficient,
interoperable, composable, and extensible software systems---can serve
arbitrary requirements for a well-defined set of entities. For example, a
workflow manager can provide methods for DAG traversing, independent of how
and when that DAG is specified or where the tasks of the workflow will be
executed. Analogously, a pilot manager can provide multi-staging and task
execution capabilities, independent of the task scheduler or the compute
resources on which tasks will be executed.

Modularity is not a design principle strong enough to realize this type of
software systems. Modularity needs to be augmented by API and coordination
agnosticism alongside an explicit understanding of the entities that define
the domain of utilization of the software system. Each system developed
following this approach, implements a well-defined set of functionalities
specific to a set of entities, with minimal assumptions about the system that
will use these functionalities or the environment in which they will be used.
Without these elements, systems developed by independent teams and not
specifically designed to work together, may require major re-engineering to
coordinate and aggregate their functionalities.

Systems like Celery, Dask, Kafka, or Docker are early examples of software
designed by implicitly following the proposed building blocks approach. These
tools implement specific capabilities like queuing, scheduling, streaming, or
virtualization for the domain of distributed computing. Consistently, they
assume a set of core entities like workloads, tasks, pipelines, or messages,
each with well-defined properties like concurrency and states. Their
integration in multiple domains shows the potential of their underlying
design approach.

This paper offers three main contributions: (i) showing the relevance of the
building blocks approach for supporting the workflows of various scientific
domains on HPC platforms; (ii) illustrating building blocks that enable
multiple points of integration, resulting in design flexibility and
functional extensibility, and providing a level of ``unification'' in the
conceptual reasoning (e.g., execution paradigm) across otherwise different
tools and systems; and (iii) showing how these building blocks have been used
to develop and integrate workflow systems for HPC platforms.

Sec.~\ref{sec:casestudies} highlights the practical impact of the building
blocks approach. All the integrations required minimal development, mainly
focused on translation layers and glue interfaces. Importantly, no
refactoring was required within the systems we integrated. Explicit and
agreed upon engineering processes was necessary to enable the integration
among systems developed by independent teams and institutions. GitHub proved
to be fundamental to enable these processes and to manage the engineering
process. Explicit agreement on written use case and software requirements
specifications greatly increase development coordination and, ultimately,
efficiency. Lastly, weekly meeting among the lead developers helped the
coding process and establishing a shared development culture.

The building blocks approach spawns many new questions. A prominent one
pertains to the issue of how we might model workflows systems and tools, so
as to provide a common vocabulary, reasoning and comparative framework.
Ref.~\cite{turilli2018comprehensive} provided the architectural paradigm for
pilot systems, however it is still unclear how an analogous paradigm would
complement the work done on reference architectures for workflow
systems~\cite{lin2009reference,grefen1998reference}, and whether, given the
very broad diversity of workflow systems and tools, we can even formulate a
single architectural paradigm. This paradigm has been elusive so far, but it
might be more fruitful to formulate system-level paradigms that have the
properties of building blocks.

It is important to outline what this paper does not attempt to achieve. This
paper presents a preliminary study focused on one approach to building blocks
for workflows systems, without a quantitative analysis of its benefits. It
does not provide qualitative insight or identify either the set of
applications or systems where a building blocks approach will surpass
alternative approaches. Finally, this paper does not discuss best practices
in the design, granularity and provision of building blocks. These are all
topics of ongoing investigation. Although preliminary, this work is not
premature: Conceptual formalisms that are too far ahead of proof-of-concepts
and demonstrable advantages are unlikely to yield practical advances. Thus,
even though the building blocks approach is still a work in progress, we
believe early reports of success are necessary.

An end-goal and intended outcome of this paper is to begin a discussion on
how the scientific workflows community---end-users, workflow designers,
workflow systems developers and HPC facilities providers---can better
coordinate, cooperate, and reduce redundant and unsustainable efforts. We
believe the building blocks approach enables an examination and investigation
of design principles and architectural patterns for workflow systems that may
facilitate this discussion.

\section{Acknowledgments}

We thank our collaborating teams led by: Peter Coveney (HTBAC); Kaushik De,
Jack Wells and Alexei Klimentov (ATLAS Project/PanDA); Charlie Laughton and
Cecilia Clementi (ExTASY); Heather Lynch (ICEBERG). We thank Daniel Smith,
Levi Naden and Sam Ellis (Molecular Sciences Software Institute (MolSSI)) for
useful discussions and insight. This work was was supported primarily by NSF
ACI 1440677 and DOE ASCR DE-SC0016280. We acknowledge access to computational
facilities: XSEDE resources (TG-MCB090174) and Blue Waters (NSF-1713749).

\bibliographystyle{abbrv}
\bibliography{CiSE-2018-11-0176.R2_Turilli}

\end{document}